\documentstyle[aps,amsmath,amssymb,psfig]{revtex}
\draft
\begin{document}

\title{Slow Light in Doppler Broadened Two level Systems}
\author{ G. S. Agarwal and Tarak Nath Dey }
\address{Physical Research Laboratory, Navrangpura, Ahmedabad-380 009, India}
\date{\today}
\maketitle
\begin{abstract}
We show that the propagation of light in a Doppler broadened
medium can be slowed down considerably eventhough such medium
exhibits very flat dispersion. The slowing down is achieved by the
application of a saturating counter propagating beam that produces
a hole in the inhomogeneous line shape. In atomic vapors, we
calculate group indices of the order of $10^{3}$. The calculations
include all coherence effects.
\end{abstract}

It is now well understood that slow light can be produced by using
the electromagnetically induced transparency
(EIT)\cite{Harris_1,Harris_2}. Many experiments have been reported
in a variety of atomic and condensed media
\cite{Kasapi,Hau,Kash,Budker,Schmidt,Turukhin}. Such experiments
reveal that the group velocity of the light pulses depends on the
parameters of the control field, which produces EIT. Various
applications of slow light have been proposed and
realised\cite{Liu,Phillips,Zibrov,Matsko,Mewes,Tarak}. Recently,
Bigelow {\it et al.} \cite{Bigelow} showed that one can produce
slow light in systems like Ruby, without the need for applying a
control field. They made a hole in homogeneous line in systems,
where the transverse and longitudinal relaxation times are of very
different order.

In this paper, we consider the possibility of producing slow light
in a Doppler broadened system. This is somewhat counterintuitive
as one would think that Doppler broadening would make the
dispersion, or more precisely, the derivative of the
susceptibility, rather negligible. We, however, suggest the use of
the method of saturation absorption
spectroscopy\cite{Hansch,Haroche,Couillaud,Sargent,Pappas} to
produce a hole of the order of the homogeneous width in the
Doppler broadened line. The application of a counter propagating
saturated beam can result in considerable reduction in absorption,
and adequate normal dispersion to produce slow light. We calculate
group index of the order of $10^3$ . We illustrate our results
using the case of the atomic vapors. However, similar or even more
remarkable results on slowing of light can be obtained for
inhomogeneously broadened solid state systems, where the densities
are large.

Consider the geometry as shown in the Fig. 1. Here a modulated
pulse of  light propagates in the direction $\hat{z}$ in a medium
of two level atoms. For simplicity we consider the incident pulse
of the form
\begin{equation}
\vec{E} (t) \equiv {\vec{\cal E}} (1+{\textrm m}\cos \nu t)
e^{i(kz-\omega t)} + c.c.,~~ k=\frac{\omega}{c}
\end{equation}
Here m and $\nu$ are the modulation index and frequency
respectively. A counter propagating cw pump field, $\vec{E}_c(t)$,
is used for producing saturation
\begin{equation}
\vec{E}_c (t) \equiv {\vec{\cal E}}_c e^{i(kz-\omega_c t)} + c.c.
\end{equation}
The effective linear susceptibility $\chi(\omega)$ of the two
level atomic systems which is interacting with the field
$\vec{\cal E} e^{i(kz-\omega t)}$ and $\vec{E}_c(t)$,  can be
calculated to all orders in the counter propagating field (2). The
effective susceptibility $\chi(\omega)$ is well known from the
work of Mollow\cite{Mollow}
\begin{eqnarray}
 \chi&=&-\frac{N|d|^2}{\hbar}\frac{1+\Delta^2{T_2}^2}
 {(1+\Delta^2{T_2}^2+4|G|^2T_1T_2)(\Delta+\delta+i/T_2)}\times\nonumber\\
 &&\left[1-\frac{2|G|^2(\Delta-i/T_2)^{-1}(\delta+2i/T_2)(\delta-\Delta+i/T_2)}
 {(\delta+i/T_1)(\delta+\Delta+i/T_2)(\delta-\Delta+i/T_2)-4|G|^2(\delta+2i/T_2)}\right],
\end{eqnarray}
where $\Delta=\omega_c-\omega_{1g}$ and $\delta=\omega - \omega_c$
are represents the detuning of the pump and probe field
respectively. For an atom moving with velocity $\vec{v}$, we
replace $\omega_c $ by $(\omega_c + kv)$. and $\omega$ by $(\omega
- kv)$. The Rabi frequency of the pump is given in terms of the
dipole moment matrix element, $\vec{d}_{1g}$, by
\begin{equation}
2G=\frac{2\vec{d}_{1g}\cdot\vec{\cal E}_{c}}{\hbar}
\end{equation}
The $T_1$ and $T_2$ are, respectively, the longitudinal and
transverse relaxation times and $N$ is the density of atoms. The
susceptibility (3) is to be averaged over the Doppler distribution
of velocities
\begin{equation}
P(kv)d(kv)=\frac{1}{\sqrt{2\pi
D^2}}e^{^{\left[-(kv)^2/2D^2\right]}}d(kv),
\end{equation}
where D is the Doppler width defined by
\begin{equation}
D=\sqrt{K_BT{\omega}^2/Mc^2}.
\end{equation}
We denote the average of $\chi(\omega)$ by ${\cal
S\texttt{}}(\omega)$. For small modulations, we can use the
approximation

\begin{equation}
S(\omega\pm\nu)=S(\omega)\pm\nu\frac{\partial S}{\partial \omega}
\end{equation}
The probe field in Eq.(1)  at the output face $z=l$ of the medium,
can be expressed as
\begin{equation}
\vec{E}(l,t)=\vec{\cal E}(1+m\cos[\nu(t+\theta)])e^{i(kl-\omega
t)+i\frac{\omega}{c}2\pi l S(\omega)}+c.c.,
\end{equation}
where the delay time, $\theta$, is defined by
\begin{equation}
\theta=2\pi l\frac{\omega}{c}\frac{\partial Re[S]}{\partial
\omega}.
\end{equation}
Note that $\theta$ will be positive if $\partial Re[S]/\partial
\omega>0$, i.e, if the medium exhibits normal dispersion. Note
further the relation of the parameter $\theta$ to the group
velocity and the group index :
\begin{equation}
v_g=\frac{c}{n_g}=\frac{c}{\left(1+2\pi Re[S(\omega)] + 2\pi\omega
\frac{\partial Re [S]}{\partial \omega}\right)}
\end{equation}
Further the imaginary part of $S$ will give the overall
attenuation of the pulse.

We present numerical results for the group index by evaluating
(10) for different intensities of the counter propagating beam. We
use typical parameter for $^{87}$Rb transition :
$T_1=T_2/2=1/2\gamma$, $\gamma=3\pi\times 10^{6}$ rad/sec,
D$=1.33\times 10^{9}$ rad/sec ( at room temperature ),
N$=2\times10^{11}$ atom/cc. We show in Fig. 2, the behavior of
real and imaginary parts of the susceptibility, $S(\omega)$,
assuming that the counter propagating pump is in resonance with
atomic transition i.e, $\omega_c=\omega_{1g}$. The imaginary part
of $S(\omega)$ shows the typical Lamb dip\cite{Demtroder} which
becomes deeper with the increase in the intensity of the
saturating beam. The real part of $S(\omega)$ exhibits normal
dispersion, which in fact, is very pronounced. It is this sharp
dispersion which can produce slow light. The calculated group
index, $n_g$, as a function of the detuning of the probe from the
atomic transition is shown in the Fig. 3. We show the behavior in
the region of Lamb dip. Clearly the group index increases with the
intensity of the saturating pump. One can calculate $n_g$ as a
function of G, for $\delta=0$, and the result is shown in the Fig.
4. To confirm these results, we also studied the propagation of a
Gaussian pulse with an envelope given by
\begin{eqnarray}
{\cal E}(t-L/c)&=&\frac{{\cal E}_0}{2\pi}\exp{\left[-(t-L/c)^2/\tau^2\right]}\nonumber\\
{\cal E}(\omega)&=&\frac{{\cal E}_0}{\sqrt{\pi\Gamma^2}}
\exp{\left[-(\omega-\omega_0)^2/\Gamma^2\right]},
\end{eqnarray}
where $\Gamma\tau$ is equal to 2 and $L$ is the length of the
medium. We use $\Gamma = 120$ kHz for our numerical simulation.
The pulse delay of $0.05~\mu sec$ due to the medium is seen in the
Figure 5. The group velocity of the pulse, calculated from the
relative delay between the reference pulse and the output pulse,
is in good agreement, with the value of group index
$[(c/v_g)=1500]$. We get the transmission of Gaussian pulse of the
order of $2.1\%$\cite{foot}. This value of transmission can be
understood by evaluating $\textrm{Im}[4\pi l\omega
S(\omega)/c]$(cf. Eq.(8)) which is found to be 3.84. This implies
a transmission $e^{-3.84}\sim 2.1\%$. The condition for
distortionless pulse propagation is that spectral width of the
Gaussian pulse to be well contained within the region of Lamb Dip
of the medium. If the pulse spectrum becomes too broad relative to
width of the Lamb dip then simple expression like (10) does not
hold. One can, however still calculate numerically the output
pulse.

In conclusion, we have shown how Lamb dip and saturated absorption
spectroscopy can be used to produce slow light with group indices
of the order of $10^3$  in a Doppler broadened medium, which
otherwise has very flat dispersion.

\begin{figure}
\caption{(a) A block diagram where the pump  $(\omega_c)$ and
probe $(\omega)$  field  are counter propagating inside the
medium. (b) Schematic representation of a two level atomic system
with ground state $|g\rangle$ and excitated state $|e_1\rangle$.}
\end{figure}

\begin{figure}
\caption{(a) and (b) The imaginary and real parts of
susceptibility $S(\omega)$ at the probe frequency $\omega$ in the
presence of pump field G. Here we considered the pump field is in
resonance. The inset shows a zoom part of the same. The common
parameters of the above four curve for $^{87}$Rb vapor are chosen
as: Doppler width parameter D$=1.33\times 10^9$ rad/sec, density
N$=2\times10^{11}$ atoms/cc, $\gamma=3\pi \times10^6$ rad/sec.}
\end{figure}

\begin{figure}
\caption{The variation of group index with the detuning of the
probe field. The parameters are chosen as : N=$2\times10^{11}$
atoms/cc, D=$1.33\times 10^{9}$ rad/sec, $\gamma=3\pi \times 10^6
$ rad/sec and $ \Delta = 0 $.}
\end{figure}

\begin{figure}
\caption{Group index variation with the Rabi frequency of the
saturating field. The parameters are chosen as :
N=$2\times10^{11}$ atoms/cc, D=$1.33\times 10^{9}$ rad/sec,
$\gamma=3\pi \times 10^6 $ rad/sec, $\Delta = 0$, and $\delta=0$.}
\end{figure}

\begin{figure}
\caption{The Solid curve shows light pulse propagating at speed
$c$ through 1 cm of vacuum. The dotted curve shows same light
pulse propagation through a medium of length 1 cm with time delay
.05$\mu$sec in the presence of saturating pump with Rabi frequency
$G=0.4\gamma$. The common parameters of the above graph for
$^{87}$Rb vapor are chosen as $N=2\times10^{11}$ atoms/cc, D$
=1.33 \times 10^{9} $ rad/sec, $ \gamma=3\pi \times 10^6 $
rad/sec, $\Delta = 0 $ and $ \delta=0 $. The transmission
intensity is $2.1\%$. The inset shows the close up of the Gaussian
pulse with a spectral width 120 kHz.}
\end{figure}

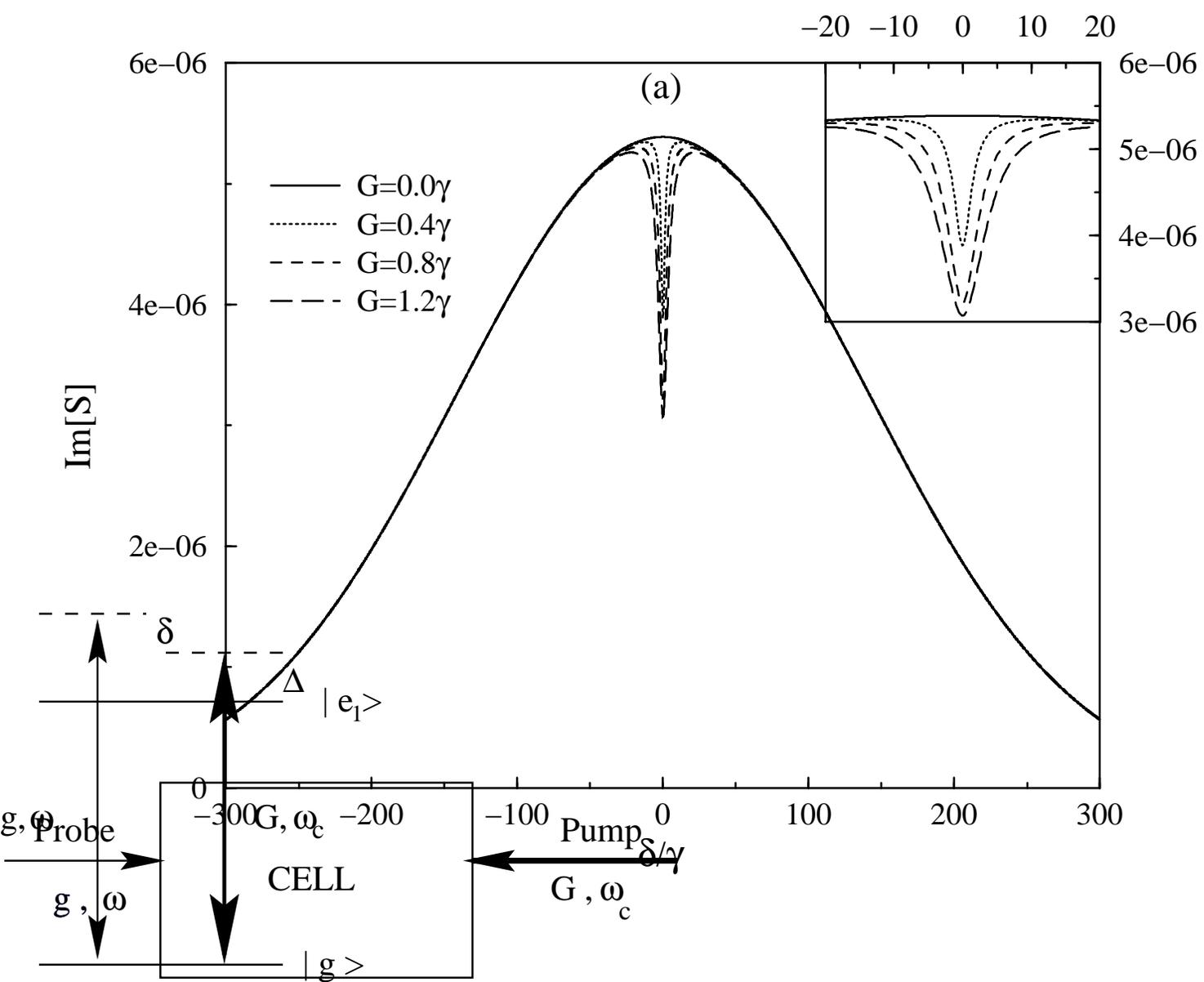
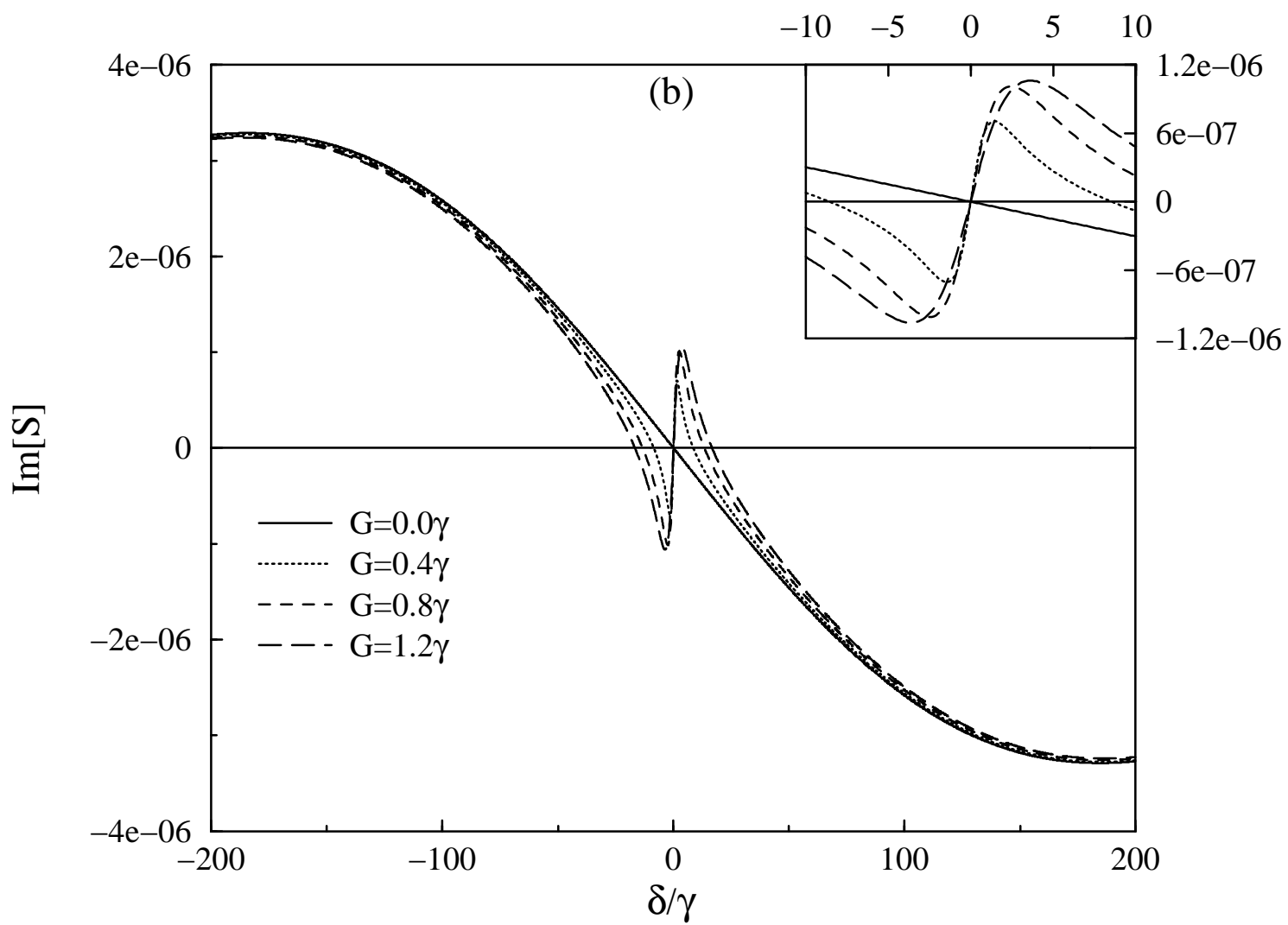
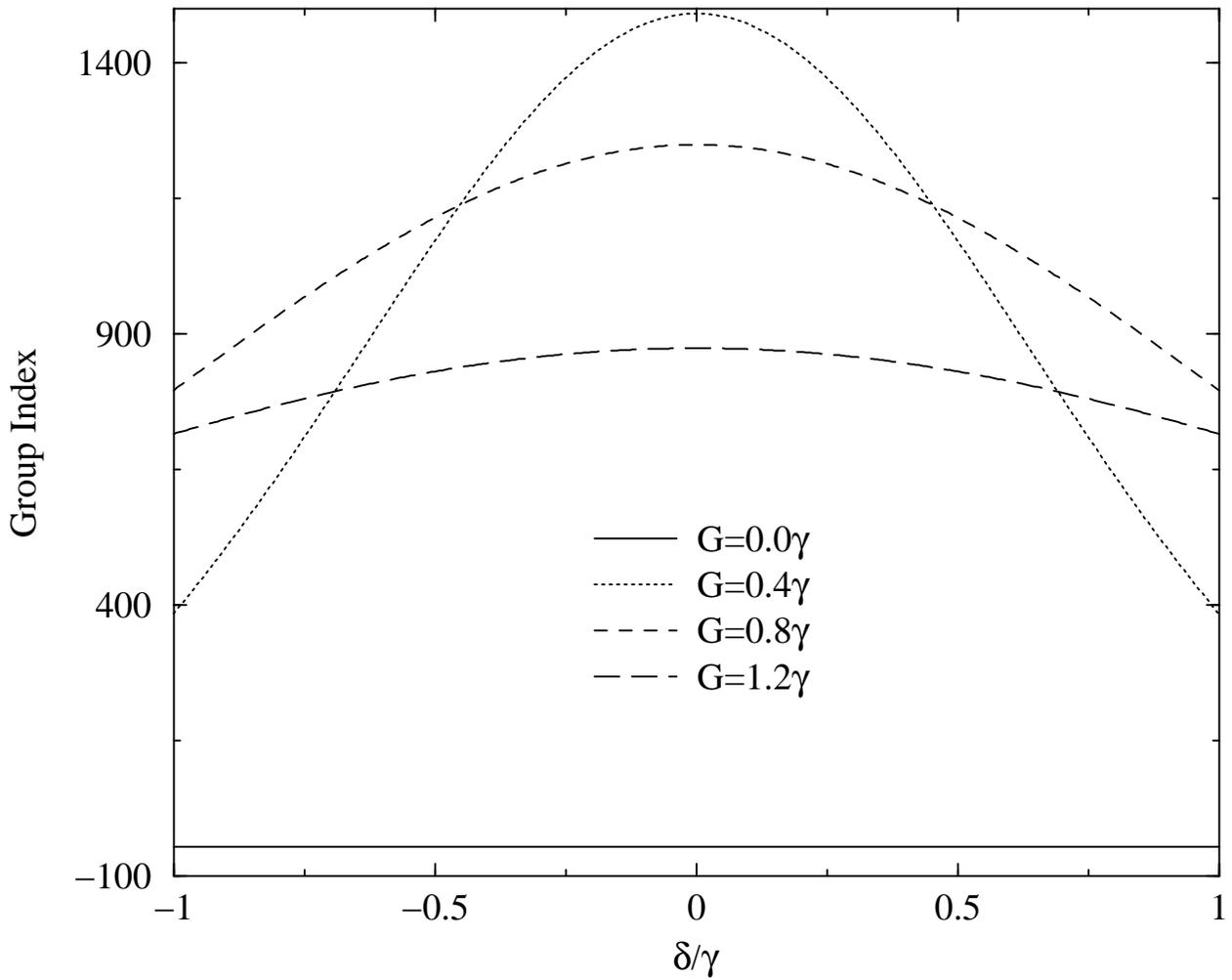
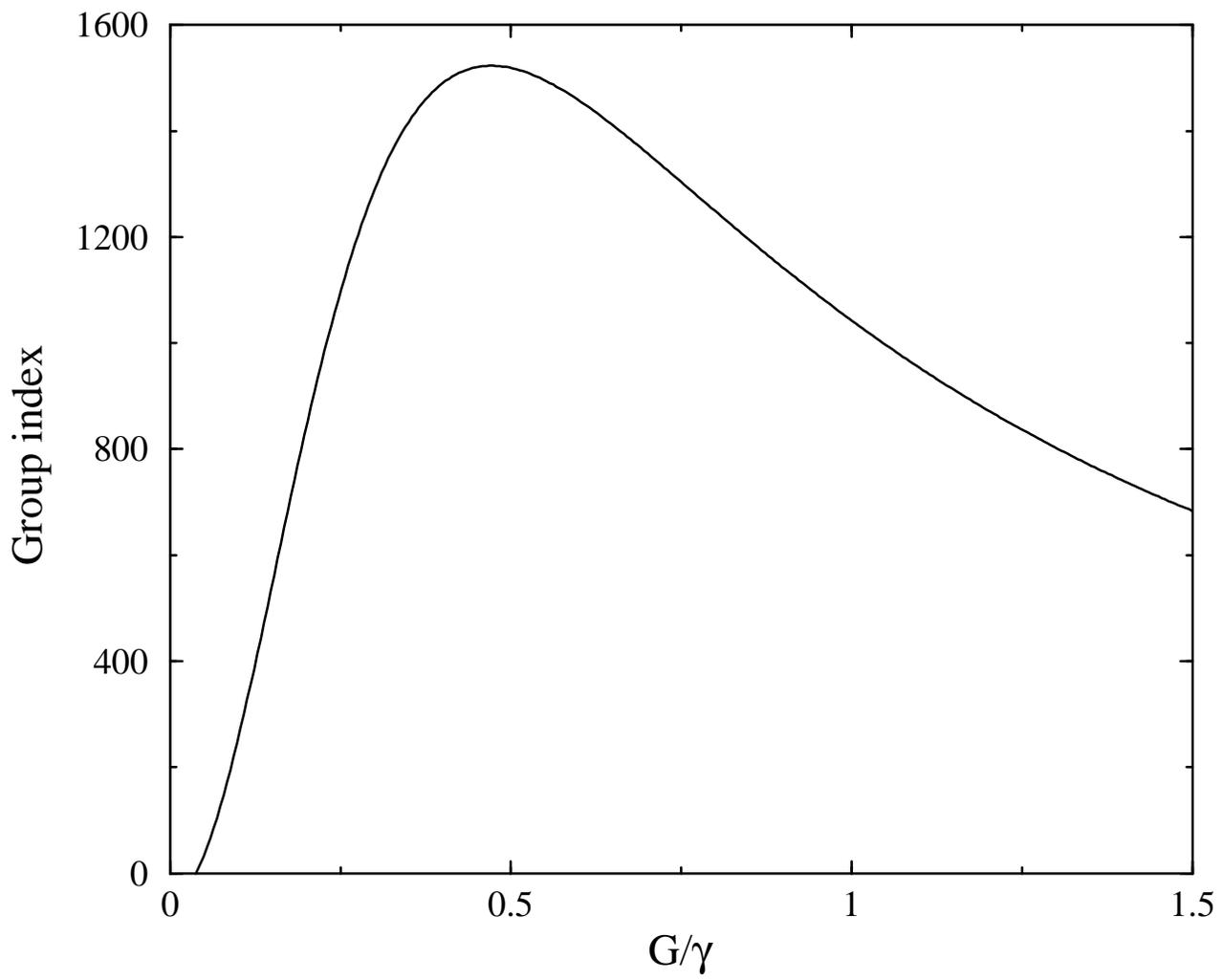
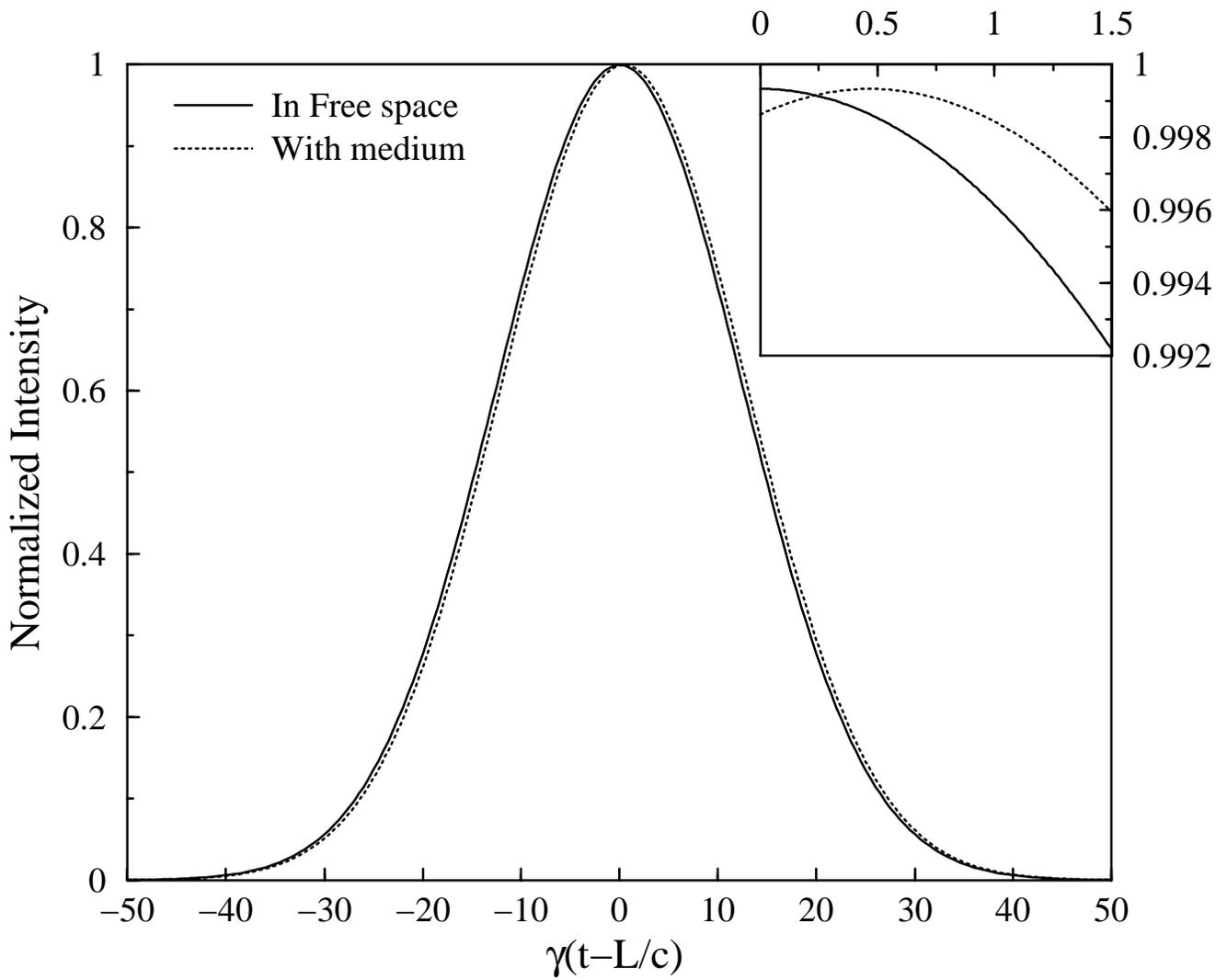
\end{document}